\documentclass[letterpaper,english,reprint,superscriptaddress,longbibliography]{revtex4-1}
\usepackage{lmodern}

\usepackage[T1]{fontenc}
\usepackage[utf8]{inputenc}
\usepackage{babel}
\usepackage{textcomp}
\usepackage{amsmath}
\usepackage{amssymb}
\usepackage{graphicx}
\usepackage[unicode=true,pdfusetitle,
 bookmarks=false,
 breaklinks=false,pdfborder={0 0 0},pdfborderstyle={},backref=false,colorlinks=false]
 {hyperref}

\makeatletter

\pdfpageheight\paperheight
\pdfpagewidth\paperwidth

\makeatother

\begin{document}
\title{Open-access microcavities: high stability without dedicated mechanical
low-pass filter in closed-cycle cryostats}
\author{M. Fisicaro}
\email{fisicaro@physics.leidenuniv.nl}

\affiliation{Huygens-Kamerlingh Onnes Laboratory, Leiden University, P.O. Box 9504,
2300 RA Leiden, The Netherlands}
\author{M.J. Rost}
\affiliation{Huygens-Kamerlingh Onnes Laboratory, Leiden University, P.O. Box 9504,
2300 RA Leiden, The Netherlands}
\author{M. Witlox}
\affiliation{Huygens-Kamerlingh Onnes Laboratory, Leiden University, P.O. Box 9504,
2300 RA Leiden, The Netherlands}
\author{H. van der Meer}
\affiliation{Huygens-Kamerlingh Onnes Laboratory, Leiden University, P.O. Box 9504,
2300 RA Leiden, The Netherlands}
\author{W. Löffler}
\email{loeffler@physics.leidenuniv.nl}

\affiliation{Huygens-Kamerlingh Onnes Laboratory, Leiden University, P.O. Box 9504,
2300 RA Leiden, The Netherlands}
\begin{abstract}
Open-access optical microcavities are Fabry-Perot type cavities consisting
of two micrometer-size mirrors, separated by an air (or vacuum) gap
typically of a few micrometers. Compared to integrated microcavities,
this configuration is more flexible as the relative position of the
two mirrors can be tuned, allowing to change on demand parameters
such as cavity length and mode volume, and to select specific transverse
cavity modes. These advantages come at the expense of mechanical stability
of the cavity itself, which is particularly relevant in noisy closed-cycle
cryostats. Here we show an open-access optical microcavity based on
scanning-probe microscope design principles. When operated at 4 K
in a tabletop optical closed-cycle cryostat without any dedicated
low-pass filter, we obtain stabilities of 5.7 and 10.6 pm rms in the
quiet and full period of the cryocooler cycle, respectively. Our device
has free-space optical access, essential for instance for full polarization
control.
\end{abstract}
\maketitle

\section{Introduction}

Optical microcavities are a powerful tool to enhance the interaction
between light and quantum systems \citep{vahalaOpticalMicrocavities2003}.
Depending on the strength of this interaction, it enables different
applications: in the weak coupling regime, the Purcell effect enables
highly efficient extraction of photons for single photon sources as
well as counteracting dephasing by the increased spontaneous emission
rate in the desired optical mode \citep{dingOnDemandSinglePhotons2016,somaschiNearoptimalSinglephotonSources2016,tommBrightFastSource2021a,thomasBrightPolarizedSinglePhoton2021,snijdersFiberCoupledCavityQEDSource2018},
while the strong coupling regime enables deterministic interaction
between distant quantum emitters and two-photon quantum gates \citep{evansPhotonmediatedinteractions2018,kimbleQuantumInternet2008,ciracQuantumStateTransfer1997,northupQuantumInformationTransfer2014,bonatoCNOTBellstateAnalysis2010,hackerPhotonPhotonQuantum2016}.
Open-access microcavities are a miniaturized version of Fabry-Perot
optical cavities, consisting of two mirrors where at least one has
a micrometer-scale curvature and which can be positioned with very
high precision with respect to each other \citep{trupkeMicrofabricatedHighfinesseOptical2005,barbourTunableMicrocavity2011a}.
While maintaining a small mode volume and potentially high Finesse
\citep{greuterSmallModeVolume2014,wachterSiliconMicrocavityArrays2019a,najerGatedQuantumDot2019},
they allow tuning of the resonance frequency and transverse cavity
mode \citep{koksObservationMicrocavityFine2022}, spatial positioning
of the cavity mode with respect to quantum emitters, and a high coupling
or collection efficiency \citep{barbourTunableMicrocavity2011a,maderScanningCavityMicroscope2015,tommBrightFastSource2021a}.

For the above-mentioned applications, many quantum emitters require
cooling down to few Kelvin, ideally around 4 K for self-assembled
semiconductor quantum dots, and the cavity needs to be mechanically
stable to maintain spectral overlap with an optical transition of
the quantum emitter. While helium bath cryostats can be made mechanically
very quiet, resulting in a very high mechanical stability of the cavity
\citep{greuterSmallModeVolume2014,najerGatedQuantumDot2019} with
root mean square (rms) fluctuations of the cavity length down to 4.3
pm, it would be highly advantageous to be able to use instead a tabletop
optical closed-cycle cryostat without the need for liquid helium,
enabling portability and minimizing the maintenance.

Because of the noisy nature of closed-cycle cryostats, the realization
of stable open-access microcavities is challenging. Recent developments
have been made with fiber-based microcavities by using mechanical
low pass filters in the cryostat \citep{bogdanovicDesignLowtemperatureCharacterization2017a,vadiaOpenCavityClosedCycleCryostat2021,fontanaMechanicallyStableTunable2021a,pallmannhighlystable2023},
and fluctuations of the cavity length as low as 15 pm rms have been
obtained in non contact mode, down to 0.8 pm rms when the cavity is
operated with direct contact between the two mirrors \citep{pallmannhighlystable2023}.

Here we show that for an open microcavity with free-space optical
access (i.e. where the light is externally coupled to the cavity by
propagating through free space) it is possible to achieve high stability
in an optical tabletop closed-cycle cryostat, without using a dedicated
mechanical low pass filter, by careful mechanical design of the open-cavity
device and using conventional feedback stabilization. In particular,
without direct contact between the two mirrors, we demonstrate sub-picometer
stability at room temperature, while in a tabletop closed-cycle cryostat
at 4 K we reach stabilities of 5.7 and 10.6 pm rms in the quiet and
full period of the cryocooler, respectively. 

\section*{Device Design}

Our open microcavity has a plano-concave configuration, where we
use a large flat bottom thin-film mirror and as the top concave mirror
a commercial chip containing an array of micro-mirrors with radii
of curvature ranging from 10 \textmu m to 100 \textmu m; the desired
top mirror can be selected by adjusting the external mode-matching
optics. The two mirrors need to be kept at a fixed distance by a device
that allows for nanometric alignment of the two mirrors with respect
to each other, along the three spatial directions and two angles,
while being insensitive to vibrations in order to operate in a mechanically
noisy environment. This is because the cryo-cooler used in closed-cycle
cryostats produces periodic mechanical pulses that propagate through
the cryostat, reaching the open-cavity device, and exciting its mechanical
resonances. As a result, the distance between the two mirrors is subject
to fluctuations induced by these vibrations. The consequences of these
vibrations are more or less severe depending on the reflectivity of
the mirrors, and therefore on the optical finesse of the cavity: in
order to maintain the optical cavity resonant with the light that
is coupled to it (or the quantum emitter in the cavity), the fluctuations
of the cavity length must be smaller than the FWHM (full width at
half maximum) of the cavity resonance. In our case we aim to operate
a cavity with a finesse $F\sim2500$ at a frequency of around $\lambda=935$
nm, which results in a FWHM of the cavity resonance in terms of cavity
length change of $\Delta L_{FHWM}\sim190$ pm given by:
\begin{equation}
\Delta L_{FHWM}=\frac{\lambda}{2F}.\label{eq:FWHM}
\end{equation}

The fluctuations in the cavity length must be much smaller than this
value. For example requiring the fluctuations to be at least 10 times
smaller than $\Delta L_{FHWM},$ we obtain a requirement on the stability
of 18 pm.

\begin{figure}
\includegraphics[width=1\columnwidth]{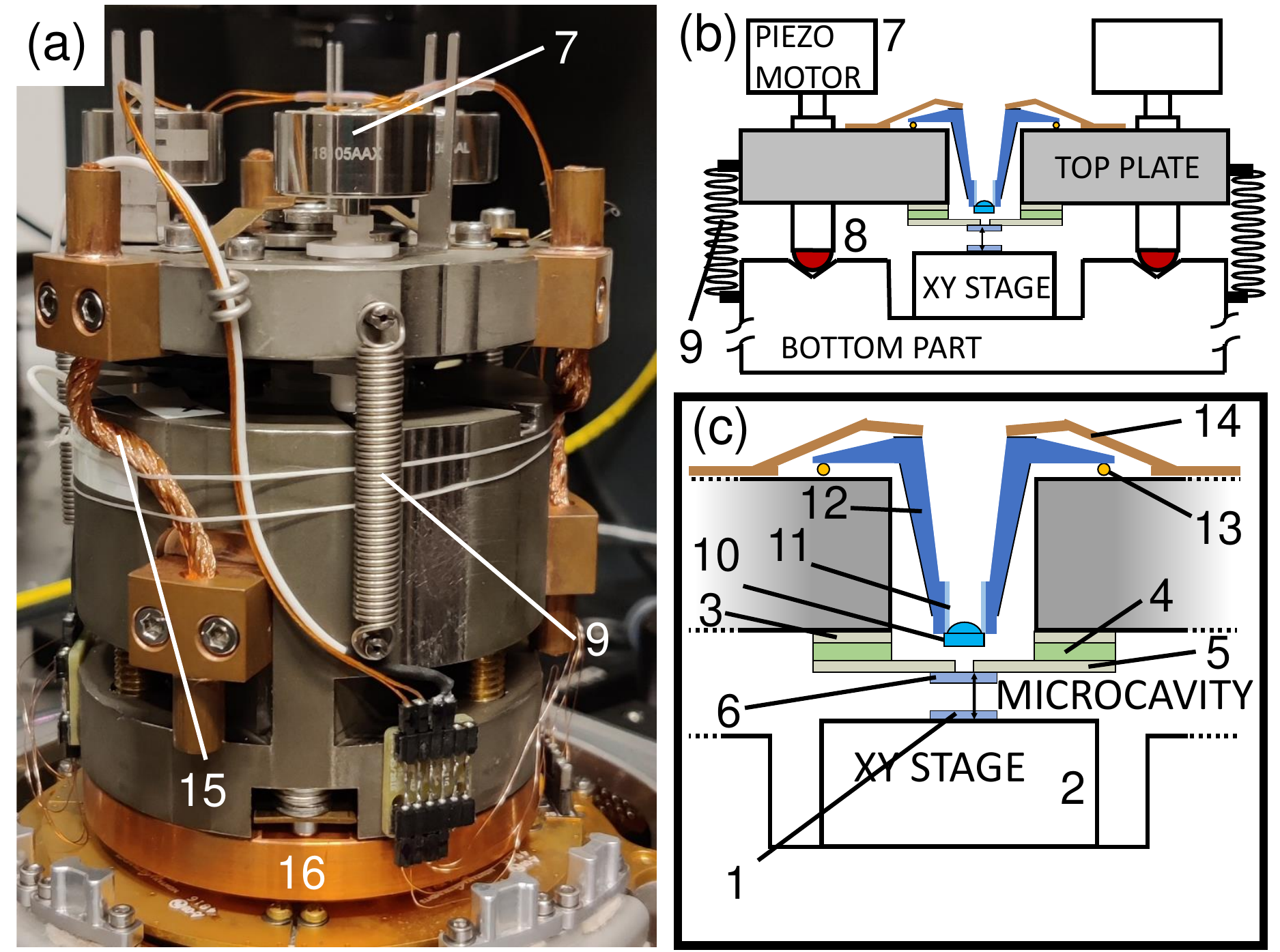}\caption{\label{Fig:cavity_mount}Picture (a) and sketch (b), (c) of the open-cavity
device. The top plate and bottom part, as well as the lens mount (11,
12), are made from invar and held together by springs; the bottom
part is clamped to the cold base plate of the cryostat by a copper
adapter disc (16). Three copper braids are used to cool the top plate
(15). The other elements are described in the main text.}
\end{figure}

This is extremely challenging, and because fully tunable open access
optical microcavities are relatively new, most of the techniques and
approaches related to the design of such system are not very well
established in the field of optics. However, the same challenges can
be found in scanning probe microscopes (SPMs) that have been developed
since the 1980s, where the challenge is to keep the tip-sample distance
stable in the same way as we want the two mirrors at a fixed distance.
In this field, ultra-high mechanical stabilities have been achieved
in a closed cycle cryostat, with a tip-sample distance variation up
to 1.5 pm \citep{hackleyHighstabilityCryogenicScanning2014} but only
if a helium exchange gas vibration isolation system is used, and even
a higher stability in a He bath cryostat, with an average vibration
level of \textasciitilde 6 fm/$\sqrt{Hz}$ \citep{battistiDefinitionDesignGuidelines2018}.
With this in mind we decided to design our device shown in Fig. \ref{Fig:cavity_mount}
following the principles and guidelines adopted in the design of scanning
tunneling microscopes (STMs) \citep{battistiDefinitionDesignGuidelines2018,chenIntroductionScanningTunneling2007}.

As shown in Fig. \ref{Fig:cavity_mount}(a) and (b), the device is
divided into two parts: the top plate, and the bottom part. The bottom
flat mirror (1) is mounted with grease on a custom-built high-stiffness
XY translation stage (2) that consists of a polished sapphire plate
on top of a XY shear piezo glued to the bottom part of the device,
similar to those used in STMs \citep{mashoffLowtemperatureHighResolution2009a}.
The XY stage has a step size in the nanometer range, with a travel
range of about 5 mm. The top plate has a hole in its center for optical
access. On the bottom side of the top plate, the top cavity mirror
mount is attached, in an assembly which consists of a ring-shaped
alumina disk (3), a ring piezo (4), another alumina ring-shaped disk
(5) and the micro mirrors array (6), all glued together withe epoxy
adhesive (EPO-TEK H74F), which is proven to deliver strong and stiff
connections in SPM vacuum applications. The thickness of the two alumina
disk has been chosen to minimize cavity length changes during cooldown.
The ring piezo (4) (Noliac NAC2125) is used for precision scanning
and active stabilization of the cavity length. The relative alignment
of the two cavity mirrors is accomplished with 3 piezo motors (JPE
CLA 2201) in a tip-tilt configuration (7), which enables adjustment
of the distance as well as the angles between the top and bottom mirrors
with nanometric precision. To ensure mechanical stability and to decouple
the movement of the top plate from a XY translation, the 3 ceramic
spherical tips (8) at the end of the spindle of the piezo motors are
resting in three v-grooves engraved on the upper side of the bottom
part. Finally, the top plate and bottom part are held together with
three metal springs (9) each exerting a force of 12 N, tuned such
that the piezo motors run. Mode matching to the microcavity mode is
achieved with an aspheric lens with 0.4 NA (10), mounted in a lens
holder consisting of an invar micrometric screw (11) in a threaded
invar cone (12) housed in the central hole of the top plate. The lens
holder has a smaller diameter than the hole in the top part, in such
a way that the lens can be translated in the XY plane for alignment
with the optical microcavity mode, and is resting on 3 ruby balls
(13) of 1 mm diameter,K glued into holes made in the top invar plate.
After aligning the lens in the XY plane and along the z direction
(with the micrometric inner screw), it is clamped down with 3 leaf
springs (14). To achieve cooling of the top plate, we thermally connected
it to the bottom part of the open-cavity device by using 3 copper
braids (15): this soft connection minimizes the amount of vibrations
transferred to the top plate. The bottom part is clamped to the cold
base plate of the cryostat by a copper adapter disk (16), therefore
even though this connection acts as a mechanical low-pas filter, there
is no additional dedicated mechanical low-pass filter between the
cryostat and our device.

The choice of the particular ring piezo mounted on the top plate
is a compromise asusually one would choose a small-diameter piezo
element with smaller capacitance and therefore higher possible bandwidth,
but our choice is limited by the mode matching lens: due to its short
working distance (3.39 mm), in order to get close to the concave top
cavity mirror, the lens must fit in the hole of the ring piezo, therefore
limiting its minimum size. The material of choice for the bottom part,
top plate and lens mount, is Invar-36 which has been proven a good
solution in designing optical cryogenic systems, due to its low thermal
expansion coefficient \citep{fedoseevRealignmentfreeCryogenicMacroscopic2022}.
The small integrated (300 K $\rightarrow$ 4 K) expansion coefficient
results in a small focal shift of the lens, and ensures minimal optical
realignment after cool-down.

The use of nanopositioners and piezo elements is crucial for the
alignment and operation of the microcavity, but the complex structure
of these actuators introduces mechanical resonance frequencies of
the device. In particular the lowest-frequency mechanical resonance
of the cavity device that significantly affects the cavity length
reduces the passive mechanical stability \citep{chenIntroductionScanningTunneling2007}
and the maximum possible bandwidth of the feedback stabilization circuit
\citep{bechhoeferFeedbackPhysicistsTutorial2005}. In order to push
this mechanical resonance as high as possible in frequency, we maximized
the stiffness of the open-cavity device by connecting the different
parts together by using either springs with high spring constants
or a high-performance epoxy adhesive. Additionally we preserved as
much as possible a rotational symmetry with respect to the central
axis of the open-cavity device: this avoids mechanical mode-splitting
and reduces the number of mechanical resonances in the device.

\section{Device characterization: feedback loop and mechanical resonances}

\begin{figure}
\includegraphics[width=1\columnwidth]{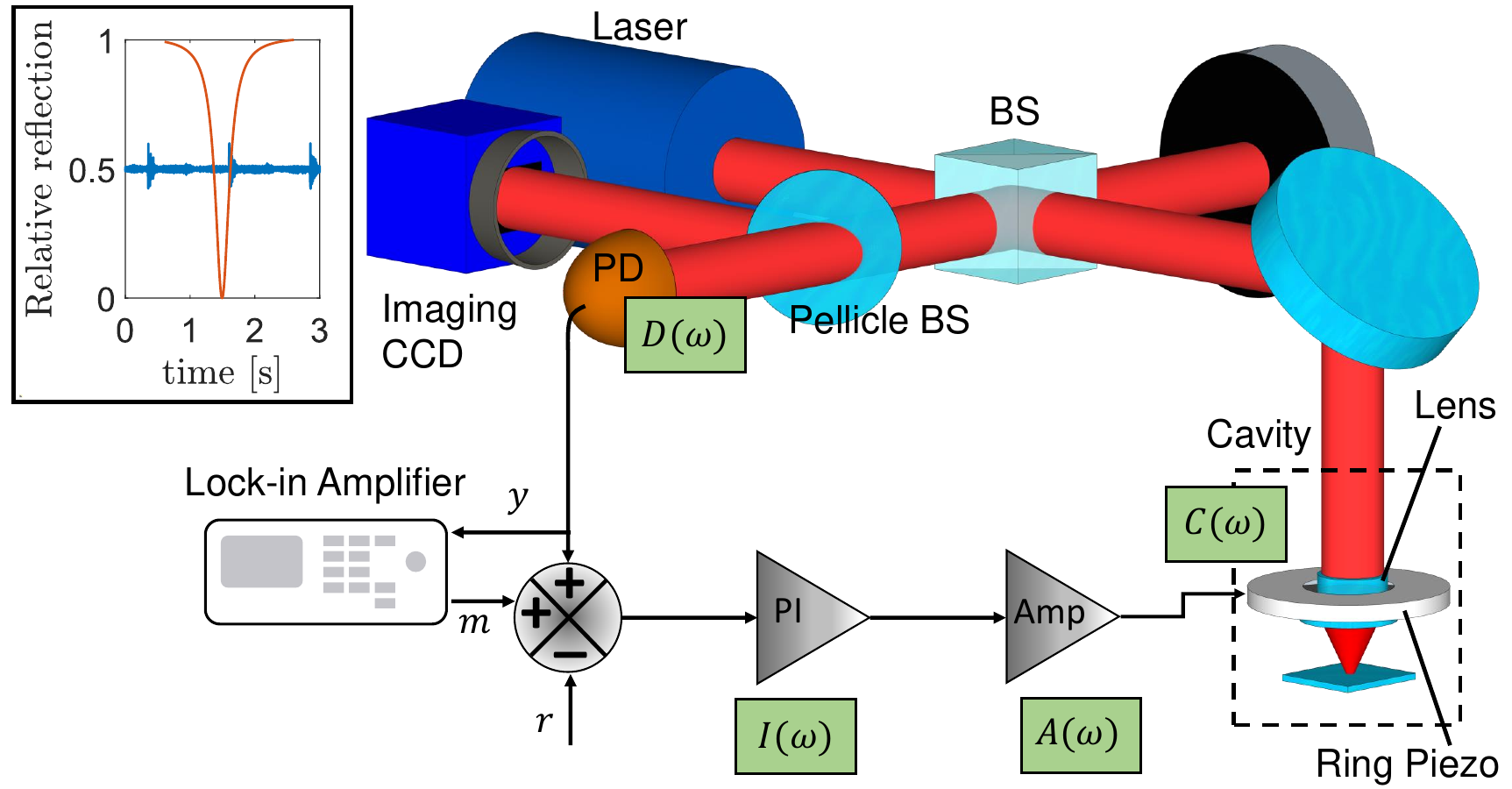}\caption{\label{fig:optical_scheme}Optical scheme for measuring the open-loop
transfer function of the feedback loop: laser light is coupled to
the cavity, and the backreflection is both monitored with an imaging
CCD and measured by a photodiode, giving a signal $y$. The difference
between $y$ and a reference signal $r$ is the error signal that
is fed into a PI controller that controls the ring piezo via an amplifier.
A lock-in amplifier is used to measure the signal $y$ from the photodiode
and to add a modulation $m$ on top of it, in order to measure the
open loop transfer function of the feedback. The inset shows the cavity
reflection dip as well as a time trace of the stabilized cavity length.}
\end{figure}
We now introduce the experimental setup that we will use in the rest
of this paper, we discuss the negative feedback circuit used for active
stabilization of the cavity length and we show how measuring the transfer
function of this feedback loop we can characterize the resonances
of the open-cavity device \citep{reinhardtSimpleDelaylimitedSideband2017,janitzHighMechanicalBandwidth2017,saavedraTunableFiberFabryPerot2021}.
This knowledge is of extreme importance not only because the lowest
mechanical resonance present in the device can pose a serious limitation
on the active stabilization, but also because it limits the passive
stability as discussed in \citep{chenIntroductionScanningTunneling2007}.
The identification of the mechanical resonances helped us improving
the stability of the open-cavity device, by redesigning some of its
parts iteratively.

\subsection{Optical setup:}

We operate the microcavity in reflection, as shown in Fig. \ref{fig:optical_scheme}:
a collimated laser beam is sent to the mode-matching lens of the open-cavity
device which couples light to the cavity, the reflected light travels
back along the same path, and is redirected to a photodiode (Thorlabs
PDA-36 EC) by a beam splitter. Before reaching the photodiode, part
of the light is also sent to an imaging CCD by a pellicle beam splitter,
allowing imaging of the micromirrors array, and also observation of
the shape of the particular transverse mode excited in the cavity.
We use narrow-linewidth tunable diode lasers in a range from $\text{\ensuremath{\lambda=\:935} nm}$
to 1000 nm. The Bragg coating of our cavity micro-mirrors is centered
at $\lambda=\:935$ nm for which the cavity has a Finesse $F\sim$
2500. The maximum coupling efficiency we obtained at $\lambda=\:935$
nm is $\eta=$ 64\%, and we reach $\eta=$ 80\% at a different wavelength
where the cavity has a finesse $F$ = 500. We have identified scattering
losses of the mirrors as the main limiting factor for the coupling
efficiency. The cavity resonance reflection dip is shown in the inset
of Fig. \ref{fig:optical_scheme}.

\subsection{Cavity stabilization feedback loop}

In order to keep the cavity resonant with the laser, we use active
stabilization of the cavity length: the signal from the photodiode
is sent through a lockbox (PI controller with scanning and locking
capabilities), then a low noise piezo amplifier (Falco Systems WMA-200)
that drives the ring piezo. Each of these stages has a specific frequency
dependent and complex transfer function, defined as the ratio between
the output and input voltages of that stage. Here and in the following
we will use interchangeably the words $`$gain' and `transfer function'.
We define the transfer functions of the PI controller as $I(\omega)$,
the piezo amplifier $A(\omega)$, the optical micro-cavity $C(\omega)$,
and the photodiode $D(\omega)$. The product of these quantities $G(\omega)=\:IACD$
is the open loop gain of the feedback loop, i.e., the ratio between
the voltage measured on the photodiode and the voltage entering the
PI controller, when the output of the photodiode is disconnected from
the PI controller, i.e., when the feedback loop is open.

We now quantify the contribution of each individual stage, in order
to provide insight into the overall gain $G$. $I$ is the gain of
the PI controller, and it is given by $I(\omega)=1/(j\omega\tau_{I})+G_{P}$,
where the frequency dependent part is given by the integrator, and
the constant part is given by the proportional term. $\tau_{I}$ is
the integration time, and inversely proportional to the integral gain.
The piezo amplifier has a gain $A(\omega)=A_{0}/(1+j\omega/\omega_{c})$
, where $\omega_{c}=\:2\pi f_{C}$. In our case $A_{0}=\:20,$ and
the cutoff frequency is $f_{c}=$ 4600 Hz when the amplifier is driving
the ring piezo which has a capacitance of 700 nF. The gain of the
cavity C($\omega)$ and the photodiode D($\omega)$ are a bit more
complicated to express, as they involve a conversion from the input
Voltage to different physical quantities. First the input voltage
is applied to the piezo, which in turn gives a displacement based
on the piezoelectric coefficient $r_{piezo}\:[\textrm{nm}/\textrm{V}]$,
then the displacement of the piezo changes the cavity length, which
in turn changes the amount of light power that is reflected back and
sent to the photodiode, as shown in the inset of Fig. \ref{fig:optical_scheme}.
The power is then converted into a voltage signal by the photodiode.
Since we will lock the cavity at half of the reflection dip (side-of-fringe
lock), which gives maximum sensitivity to changes of the cavity length,
we will give the overall gain of the cavity and photodiode together:
$CD$($\omega)=-r_{piezo}\times2\eta V_{0}F/\lambda$, where $\eta$
is the coupling efficiency of the light to the cavity, $F$ is the
finesse, $\lambda$ is the wavelength of the incident light and $V_{0}$
is the voltage on the photodiode when the cavity is tuned off-resonance.
This gain is different for different measurements, since we vary the
cavity finesse by adjusting the laser wavelength, which also changes
the laser power and with this $V_{0}$. Moreover the piezoelectric
coefficient is $r_{piezo}=3.8\textrm{ \textrm{nm}/\textrm{V}}$ at
room temperature, but becomes a factor $\sim$ 3 lower when operating
the cavity at 4 K. As an example, at room temperature we have $CD$
= 7.1 with $\lambda=\;935$ nm, $F$ = 2500 and a typical value of
$V_{0}=0.7$ V and $\eta=\:0.5$.

\subsection{Identification of mechanical resonances}

\begin{figure}
\includegraphics[width=1\columnwidth]{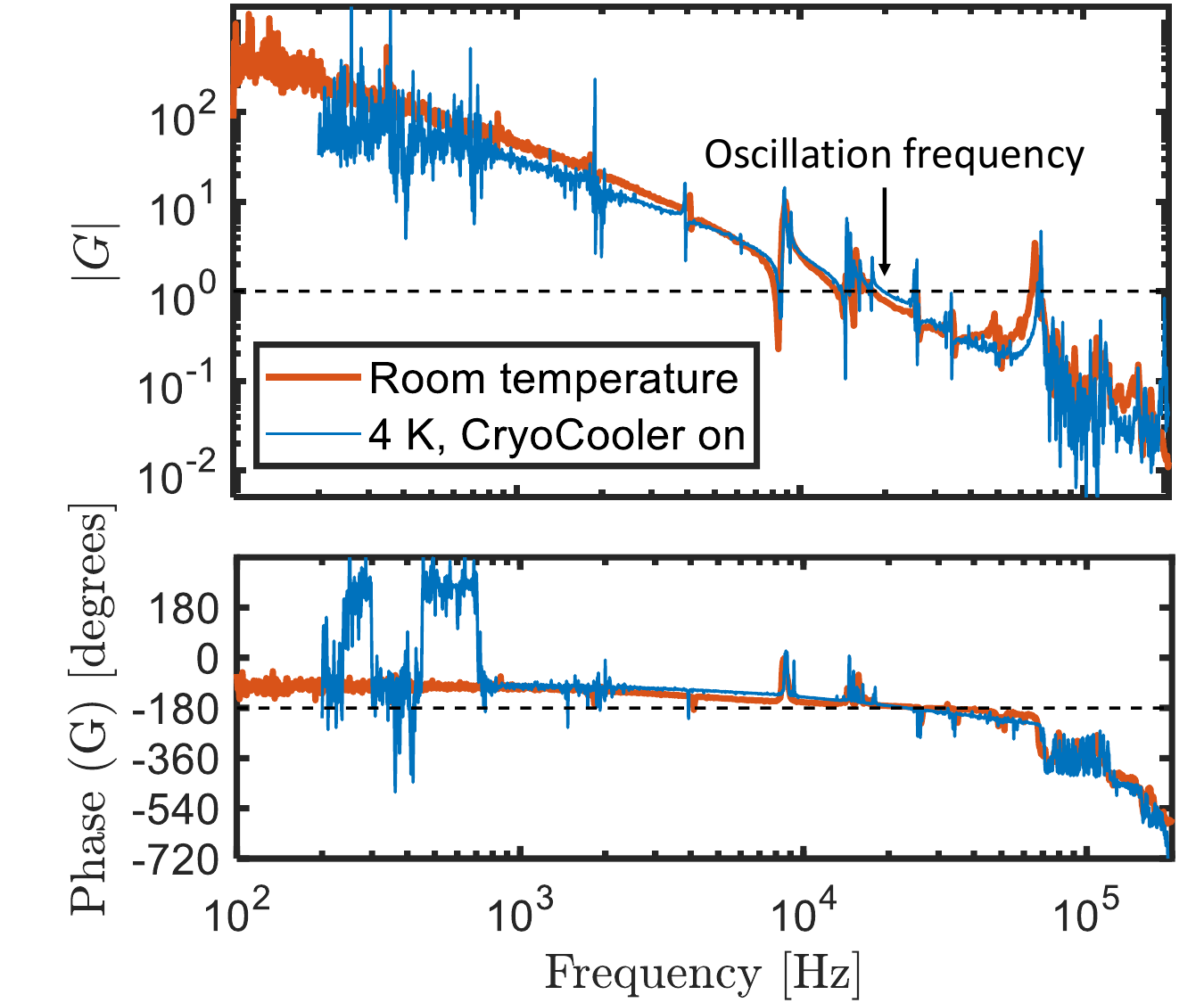}

\caption{\label{fig:optical_measured_transferFunction-1}Measured amplitude
and phase (unwrapped) of the feedback loop transfer function for two
different conditions: device mounted in the cryostat with the cryocooler
switched off at room temperature and ambient pressure (red), and at
4 K with the cryocooler on (blue). The positions of the mechanical
resonances is not significantly shifted at low temperature. The dashed
lines corresponds to unity gain and -$\pi$ phase respectively; the
feedback loop oscillates at 22.8 kHz if the gain is set too high.}
\end{figure}

Up to now $CD$ is frequency independent, but we will now show that
this is not true, and how its frequency dependency can be used to
find the mechanical resonances of the open cavity device. The ring
piezo is mechanically coupled to the rest of the open cavity device,
and therefore to all its mechanical resonances. Because of this, the
piezoelectric coefficient $r_{piezo}$ is in fact frequency dependent,
and will show a resonance–anti-resonance behavior in proximity of
the mechanical resonances present in the system \citep{ryouActiveCancellationAcoustical2017}.
This shows up in the total open loop transfer function of the feedback
loop as shown in Fig. \ref{fig:optical_measured_transferFunction-1}.
This measurement has been done with a locked cavity, using a lock-in
amplifier to apply a frequency-swept small modulation $m$ to the
input port of the PI controller and to measure the signal from the
photodiode $y$, from which the open-loop gain $G$ is calculated
as:
\begin{equation}
G(\omega)=\frac{H(\omega)}{1+H(\omega)},\!\textrm{with }H(\omega)=\frac{y(\omega)}{m}
\end{equation}

By performing several measurements of the transfer function $G$ while
damping, removing or changing the configuration of specific parts
of the open-cavity device, we were able to identify the origin of
several mechanical resonances, and to redesign the responsible components
in order to remove or shift them to higher frequencies. After device
optimization, we obtain the transfer function $G$ shown in Fig. \ref{fig:optical_measured_transferFunction-1}
of the device mounted in the cryostat at room temperature (red curve)
and at 4 K with the cryocooler on (blue curve). The cryocooler leads
to noise, particularly visible at low frequencies. We observe only
a small shift and change of the resonance frequencies while cooling
down. At 4 K the first -$\pi$ phase crossing happens at 4.1 kHz and
the first unity gain frequency is at 8.58 kHz, while the feedback
loop oscillation frequency is 22.8 kHz. For a simple system with only
one mechanical resonance these 3 frequencies would coincide, but our
open-cavity device is more complex as there are many mechanical resonances
coupled with the resonance of the ring piezo. We identified some of
the resonances to be drum modes of the top plate (8.58 kHz, 14.3 KHz,
15.6 kHz, 26 kHz and 35 kHz). We were not able to pinpoint the origin
of the resonances at 1.8 kHz, 2 kHz and 4.1 kHz.

\section{Stability at room temperature}

\begin{figure*}
\includegraphics[width=1\textwidth]{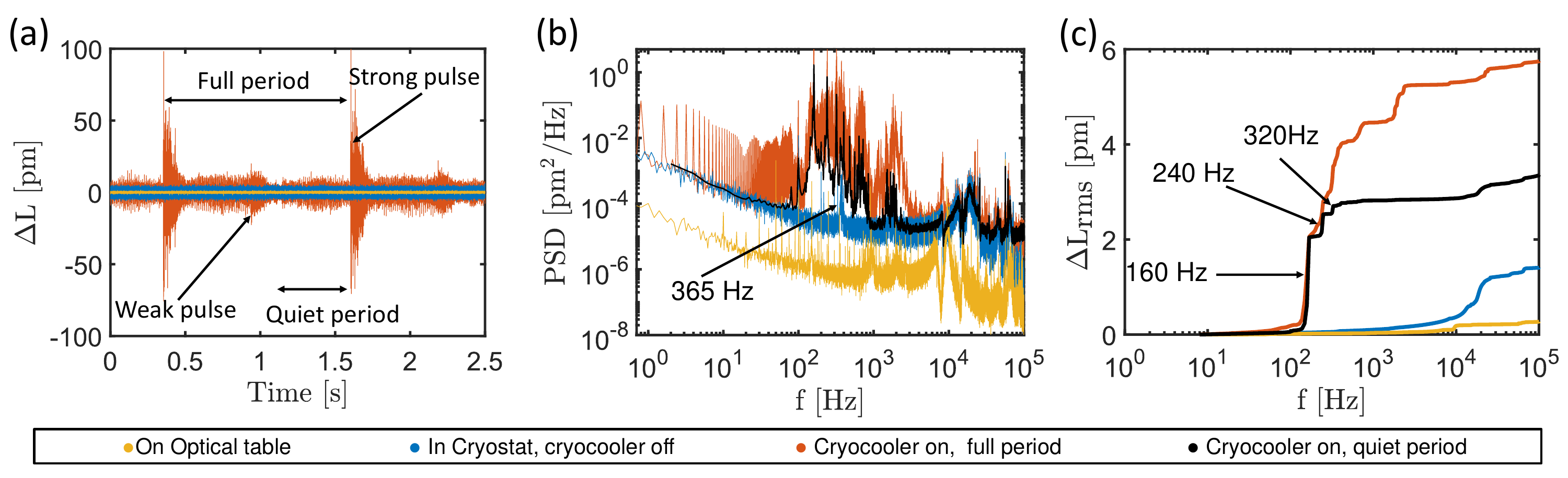}

\caption{\label{fig:Room-temperature-measurements}Room-temperature stability
measurements of the open-cavity device for 3 different conditions:
mounted on the optical table (yellow), mounted in the cryostat under
vacuum (blue), and in the cryostat under vacuum and with the cryocooler
switched on (red). Note that the measurement in yellow corresponds
to an earlier version of the cavity device. (a) Time trace of the
signal from the photodiode, for a cavity locked at half of the reflection
dip. (b) Power spectral density of the signal from the time traces.
(c) Cumulative cavity length fluctuation as a function of frequency.
The cavity finesse is $F=1500$ in all cases.}
\end{figure*}

To measure the stability of the cavity, we use the scheme shown in
Fig. \ref{fig:optical_scheme} but without the modulation signal $m$,
and again locking the cavity to the laser frequency using a side-of-fringe
lock. In particular, we lock the cavity at half the depth of the reflection
dip, and we record 10 time traces of 10 s of the photodiode signal
at a sample rate of 200 kHz. After a proper calibration (see Appendix
\ref{sec:Calibration-of-the}) we calculate the power spectral density
(PSD) of these time traces, corresponding to fluctuations in the cavity
length. In Fig. \ref{fig:Room-temperature-measurements} we show data
measured (i) on the optical table resting on air dampers, (ii) in
the cryostat under vacuum, (iii) in the cryostat under vacuum with
the cryocooler switched on but still at room temperature, (iv) same
conditions as iii but only considering the quiet periods of the cryocooler,
corresponding to 40\% of the full period. Panel (a) shows an exemplary
section of the time traces, and (b) shows the corresponding PSDs.
The finesse of the cavity is 1500, and the PI parameters are $\tau_{I}$
= 30 $\mu s$, $G_{P}$ = 0.17.
Comparing the measurements without cryocooler, it is clear that the
cavity length is more stable when the device is mounted directly on
the optical table (yellow trace), than when it is mounted in the cryostat.
The rms cavity length fluctuations integrated from 1 Hz to 100 kHz
are $\Delta L_{rms}=0.29$ pm on the optical table, and $\Delta L_{rms}=1.5$
pm mounted in the cryostat. This decrease in stability can be attributed
to mechanical noise entering the cavity through the connections of
the cryostat to the rest of the lab environment (such as the high-pressure
helium flex lines): this noise bypasses the low-pass filter of the
optical table. The small peak at 365 Hz in the PSD (blue) of Fig.
\ref{fig:Room-temperature-measurements}(b) corresponds to a mechanical
resonance in the base of the cryostat.

Focusing on the measurement with the cryocooler switched on, the time
traces show a stronger and a weaker pulse, both with period 1.25 s
(0.8 Hz) separated by 0.625 s. In the PSD we can see harmonics of
this frequency up to kHz frequencies. By calculating the PSD for different
bandwidths, we can plot the cumulative cavity length fluctuation in
Fig. \ref{fig:Room-temperature-measurements}(c), this enables best
understanding of the frequencies that contribute most to the mechanical
noise.

Because of the large cavity length changes caused by the spikes of
the cryocooler, here and in the rest of the paper we define a quiet
period in between the small spike and the big spike, corresponding
to 40\% of the time between 2 large spikes, and we will use it to
calculate the PSD and noise characteristics of the cavity device.
In this way the PSD will also be cleaner as all the cryocooler harmonics
caused by the train of pulses in the time trace will be removed, and
it will be easier to identify mechanical resonances. The PSD and the
cumulative rms noise are shown as black lines in Fig. \ref{fig:Room-temperature-measurements}(b)
and (c). During the quiet period we obtain an rms cavity length fluctuation
of $\text{\ensuremath{\Delta L_{mrs}=3.4}}$ pm, as opposed to 5.7
pm if measured over the full period of the cryocooler. As can be seen
from the cumulative cavity length fluctuations, the major contribution
to the mechanical noise comes from frequencies lower than one kHz,
in particular 160 Hz, 240 Hz, 320 Hz. These frequencies are not resonances
of the cavity device, as they are not visible in the transfer function
measurement shown in Fig. \ref{fig:optical_measured_transferFunction-1},
and they appear in the cryostat measurements only when the cryocooler
is switched on. Further evidence of this is presented in Appendix
\ref{sec:Comparison-of-cryocooler}.

\section{Stability in a closed-cycle cryostat at 4 K}

Now we describe the cool-down, re-alignment and measurements under
cryogenic conditions. With the aid of a copper adapter plate, the
open-cavity device is tightly mounted on the base of the closed-cycle
cryostat. The cavity is aligned at room temperature using the nanopositioners
and the external optics, before the system is cooled down to 4 K.
By monitoring the free spectral range of the cavity during cool-down,
we observe a change in the cavity length of $\Delta L=5.5$ \textmu m.
Once at 4 K, the cavity length can be readjusted with the 3 piezo
motors if needed. We control the finesse of the cavity by detuning
the laser frequency, which changes the mirror reflectivity, such that
we obtain the highest finesse cavity that can be locked despite the
noise from the cryocooler. The PI parameters of the feedback system
are well-controlled as described in section II. In practice, the highest
possible finesse is obtained when the peak-to-peak displacement of
the cavity length caused by the cryocooler is slightly smaller than
the FWHM of the reflection cavity resonance dip, beyond that the lock
is lost during every cryocooler cycle. We usually start with a laser
frequency far detuned from the thin-film mirror stop-band center,
locking the cavity and optimizing the PI parameters to minimize the
mechanical noise in the cavity length, then slightly changing the
laser frequency in order to increase the finesse of the cavity, and
iterating the procedure until the PI controller is not able to keep
the cavity locked.

The thin-film mirror design wavelength is 935 nm, resulting in a cavity
finesse of 2500. In our case the highest finesse we were able to achieve
is 1800 at $\lambda$ = 990 nm, for which we have a stable lock that
lasts for at least 2 days without any adjustment, by using the PI
parameters $\tau_{I}$ = 50 $\mu s$ and $G_{P}$ = $0.03$. As described
in the previous section, once the cavity is locked at half of the
reflection dip, we record 10 time traces each 10 s long of the photodiode
signal. After converting this voltage signal into cavity length fluctuations
(see Appendix \ref{sec:Calibration-of-the}), we calculate the PSD
and the cumulative integrated rms noise to identify the frequencies
that contribute most strongly.

\begin{figure}
\includegraphics[width=1\columnwidth]{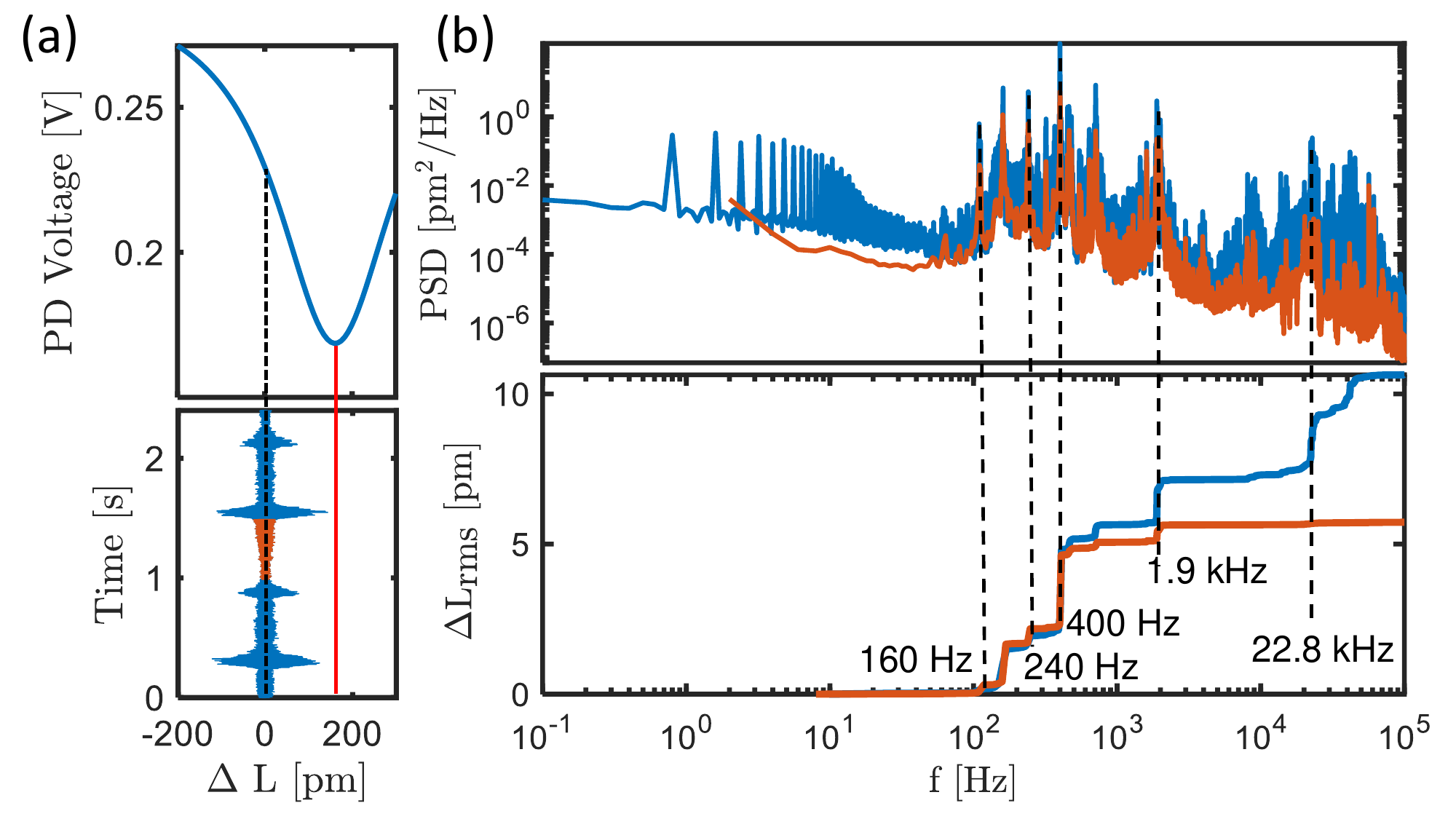}\caption{\label{fig:4K_highFinesseMeasurement}Stability measurement at 4 K
for a cavity finesse $F$ = 1800 ($\lambda=990$ nm). Panel (a) shows
the calibration curve and a section of a time trace of the cavity
length displacement. The red-highlighted part in (a) corresponds to
the quiet period of the cryocooler. In (b) are shown the PSD and cumulative
noise corresponding to the quiet period (red) and full period (blue)
of the cryocooler. The rms mechanical noise is 5.7 pm in the quiet
period, ad 10.6 pm in the full period.}
\end{figure}

In Fig. \ref{fig:4K_highFinesseMeasurement}(a) we show a reflection
dip and a small portion of the calibrated time trace taken with a
locked cavity, that represents the fluctuations of the cavity length
from the locked position. The red vertical line corresponds to the
maximum peak cavity length fluctuation that allows locking of the
cavity, while the red-highlighted part of the time trace corresponds
to the quiet period of the cryocooler (40\% of the full period). Using
the calibrated time trace, we calculate the power spectral density
and the cumulative cavity length fluctuation for the quiet period
(red) and for the full period (blue) of the cryocooler, as shown in
Fig. \ref{fig:4K_highFinesseMeasurement}(b). The total rms mechanical
noise in a 100 kHz bandwidth is 5.7 pm during the quiet period of
the cryocooler and 10.6 pm during the full period. As it is clearly
visible in the plot, also at 4 K, the major contribution to the noise
comes from the same low frequencies that were present at room temperature
with operating cryocooler, in particular at 160 Hz and 240 Hz. In
addition, now there is also mechanical noise at 400 Hz, which we think
is a mechanical resonance of the base of the cryostat that shifted
from the room temperature value of 365 Hz. Only a small part of the
noise comes from resonances of the open-cavity device itself, one
small contribution at 1.9 kHz (compare Fig. \ref{fig:optical_measured_transferFunction-1}),
and a contribution from 22.8 kHz, which is the oscillation frequency
of the feedback loop. We note that the contribution from 22.8 kHz
is significant only in the full period curve (blue), because the pulses
from the cryocooler bring the feedback close to instability.

\section{Conclusions and outlook}

In conclusions, we have developed an open-access optical microcavity
compatible with a tabletop optical closed-cycle cryostat. At 4 K, the
open cavity device has stabilities of 5.7 and 10.6 pm rms during the
quiet and full period of the cryocooler cycle, respectively. The device
allows for full nanometric but millimiter-range tunability of the
cavity along the three spatial directions and two angles, and minimal
optical realignment when cooling down to 4 K. The key to this is an
extremely high-stiffness and compact design, which allows using active
feedback stabilization of the cavity with a very high bandwidth.

Most importantly, our design does not use a dedicated mechanical low
pass filter, which would complicate integration of the cavity in a
free-space optical setup, essential for instance for full polarization
control. Nevertheless a low-pass filter would increase the stability
of our open-access optical microcavity. We estimate that a combination
of a mechanical low pass filter at around 50 Hz and electronic filtering
in the feedback loop \citep{ryouActiveCancellationAcoustical2017,okadaExtendingPiezoelectricTransducer2020},
will allow us to achieve at cryogenic temperatures the same cavity
stability as on an optical table at room temperature.

\section{Aknowledgements}

We would like to thank G. Verdoes and E. Wiegers for helping with
the construction of the open-cavity device, and K. Heeck for useful
discussions on the electronics part. We acknowledge funding from a
NWO Vrij Programma Grant (QUAKE, 680.92.18.04), the European Union’s
Horizon 2020 research and innovation programme under grant agreement
No. 862035 (QLUSTER), and the Quantum Software Consortium.

\bibliographystyle{naturemagwV1allauthors}
\nocite{*}
\bibliography{PaperLiterature_final_correct}

\renewcommand{\thefigure}{A\arabic{figure}}\setcounter{figure}{0}\renewcommand{\theequation}{A\arabic{equation}}\setcounter{equation}{0}
\appendix

\section{Comparison of cryocooler in low- and high-power mode}

\label{sec:Comparison-of-cryocooler}The Gifford-McMahon Cryocooler
used in our cryostat can be operated in a low-power (LP) and high-power
(HP) mode. We compare here the effects of the two different settings
on the stability of the open-cavity device. As a result of the exchange
Helium gas undergoing compression-expansion cycles, the cryocooler
introduces mechanical noise in the open-cavity device, in the form
of pulses. The frequency of this cycle is 0.8 Hz for LP and 1.01 Hz
for HP mode, corresponding to the frequency at which the compressed
gas enters the regenerator through the high-pressure line. In the
middle of the cycle, the Helium gas expands and returns to the compressor.
This results in two mechanical pulses per cycle of the base plate
of the cryostat.

\begin{figure}
\includegraphics[width=1\columnwidth]{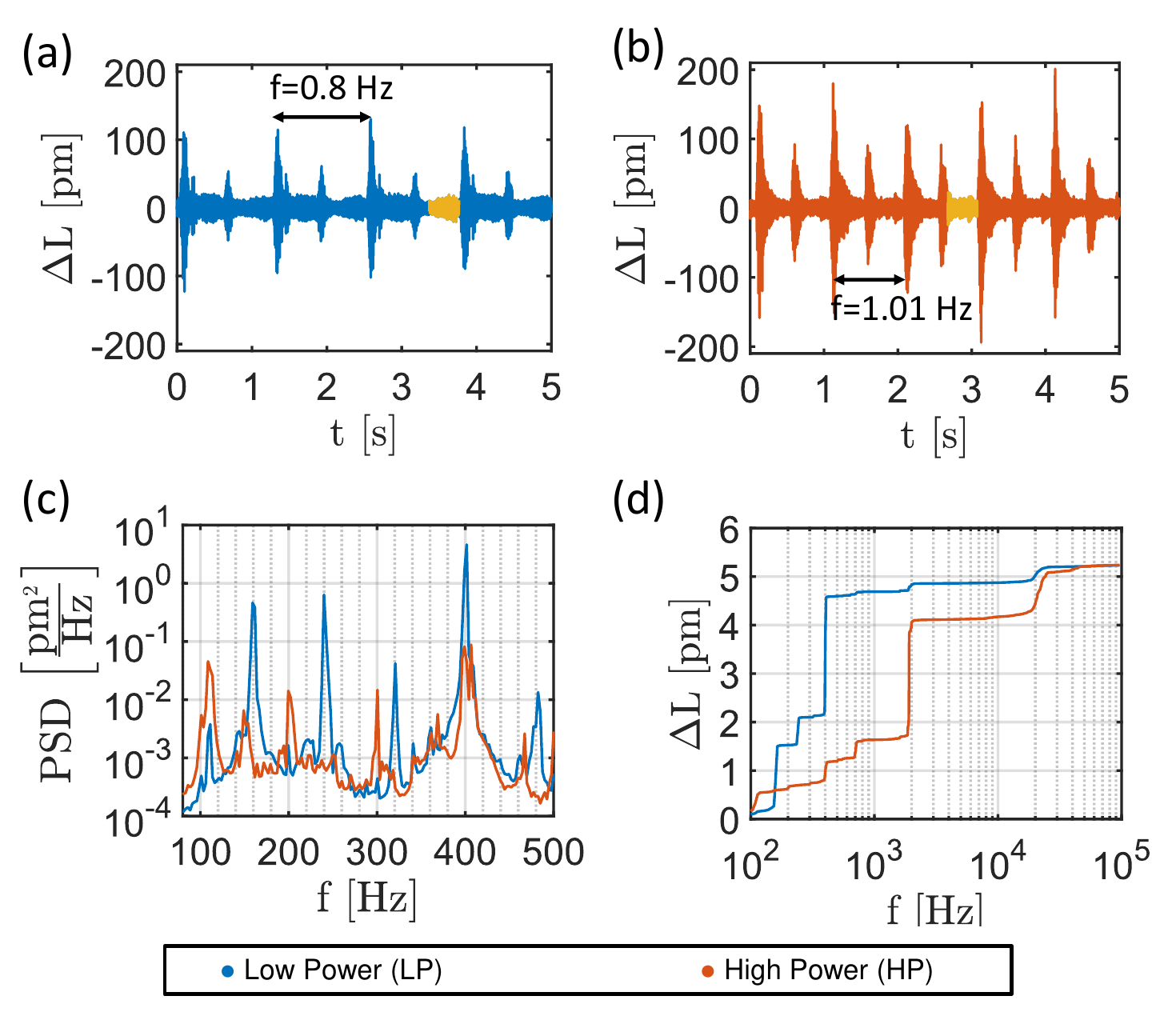}

\caption{\label{fig:LowPower_vs_HighPower}Comparison of low- and high-power
mode. Panels (a) and (b) show the measured cavity length fluctuations
as a function of time for LP (a) and HP (b) mode. The yellow portion
of the data is used to calculate the PSD shown in (c) and the cummulative
length fluctuations (d), blue curves for LP and red curves for HP
mode.}
\end{figure}

In Fig. \ref{fig:LowPower_vs_HighPower}(a) and (b) we show the fluctuation
of the cavity length measured over time with the cryocooler in the
LP and HP mode, where the cavity length is actively stabilized by
locking at half of the reflection dip as explained in the main text.
Each plot shows two kinds of pulses corresponding to the high and
low pressure part of the cycle. Comparing Fig. \ref{fig:LowPower_vs_HighPower}(b)
to Fig. \ref{fig:LowPower_vs_HighPower}(a) we see that in addition
to the higher frequency of the pulses, the peak-to-peak amplitude
of the pulse is also higher, as a result of the increased pressure
difference of the Helium. 
We calculated the power spectral density (PSD) as explained in the
main text, by considering the yellow sections of the time traces in
Fig. \ref{fig:LowPower_vs_HighPower}(a) and (b) which corresponds
to the quiet period, i.e. 40\% of the cycle for the high power setting.
We considered the same section also for the low power mode, so that
the frequency separation of the points in the PSD is the same for
both configurations, making a comparison easier. As explained in the
main text, we take 10 time traces of 10 seconds each, for each time
trace we select the orange sections in each cycle of the cryocooler,
we calculate the PSD of each section and finally we average them together.
The result of this process is shown in Fig. \ref{fig:LowPower_vs_HighPower}(c),
where we plot only a small frequency range 80-500 Hz because outside
of this range, there are not many differences between LP and HP mode,
except for the peak heights. In particular, in both PSDs we see a
400 Hz peak corresponding to a resonance of the base of the cryostat
as motivated in the main text. All other low frequency peaks that
are visible in the LP mode (160 Hz, 240 Hz and 320 Hz), are not present
in HP mode, where there are now three extra peaks at 150 Hz, 200 Hz
and 300 Hz, most likely harmonics of 50 Hz line noise.

Despite we do not know the exact origin of the peaks at 160 Hz, 240
Hz and 320 Hz, we can conclude that it is noise introduced by the
cryocooler and these peaks do not correspond to mechanical resonances
in the system because otherwise they should be visible also with the
cryocooler in HP mode. Because of the constant frequency separation
between these peaks, one hypothesis is that they originate from the
inverter used in the He compressor, as in LP mode the frequency of
the inverter is 40 Hz.

Fig. \ref{fig:LowPower_vs_HighPower}(d) shows the cumulative cavity
length fluctuation measured in the quiet period of the cryocooler
cycle. The total rms noise $\Delta L_{rms}$ = 5.3 pm is nearly the
same for the two modi, despite the frequencies that contribute to
the noise are different. We have used the same PI parameters for
both measurements, optimized for the LP mode for minimum vibrations.
The feedback loop is less stable in the HP mode due to a higher peak-to-peak
amplitude of the pulses excited by the cryocooler, as visible in the
22.8 kHz peak corresponding to the oscillation frequency of the feedback
loop.

\section{Calibration of the Photodiode signal}

\label{sec:Calibration-of-the}In order to measure the stability of
the cavity device, we lock the cavity length to half of the reflection
dip and we record time traces of the voltage signal from the photodiode,
as explained in the main text. This voltage must be multiplied by
a conversion factor in order to give a measure of the cavity length.
While this is trivial in the case of a quiet system where the cavity
length does not change much and therefore this conversion factor is
simply the slope of the reflection dip at the locking point, in the
case of a mechanically noisy system, it is more complicated. In fact,
when the cavity length changes significantly, the error signal can
not be approximated anymore as being linear, and in order to retrieve
the exact cavity length, we must take the nonlinearities into account.
For this purpose we calibrate the voltage signal of the photodiode
$V_{PD}$ as a function of the cavity length $L$, by linearly changing
the latter via the ring piezo and recording the reflection dip, which
is fitted with a Lorentzian

\begin{equation}
V_{PD}=V_{0}+\frac{A}{(L-L_{0})^{2}+(\frac{w}{2})^{2}}\ .\label{eq:Lorentzian}
\end{equation}

Inverting the equation, we are able to obtain a conversion factor
from photodiode voltage to cavity length :

\begin{equation}
L=L_{0}\pm\sqrt{\frac{A}{V_{PD}-V_{0}}-\left(\frac{w}{2}\right)^{2}}\label{eq:Lorentzian_inverted}
\end{equation}

where the $\pm$ sign is chosen depending wheter the cavity is locked
on the left or right side of the reflection dip. $A$, $V_{0},$ $L_{0}$
and $w$ are fitting parameters, $w$ is the full width at half maximum
of the reflection dip, and $L_{0}$ is the resonance cavity length,
for which the reflection dip has its minimum.

\begin{figure}
\includegraphics[width=1\columnwidth]{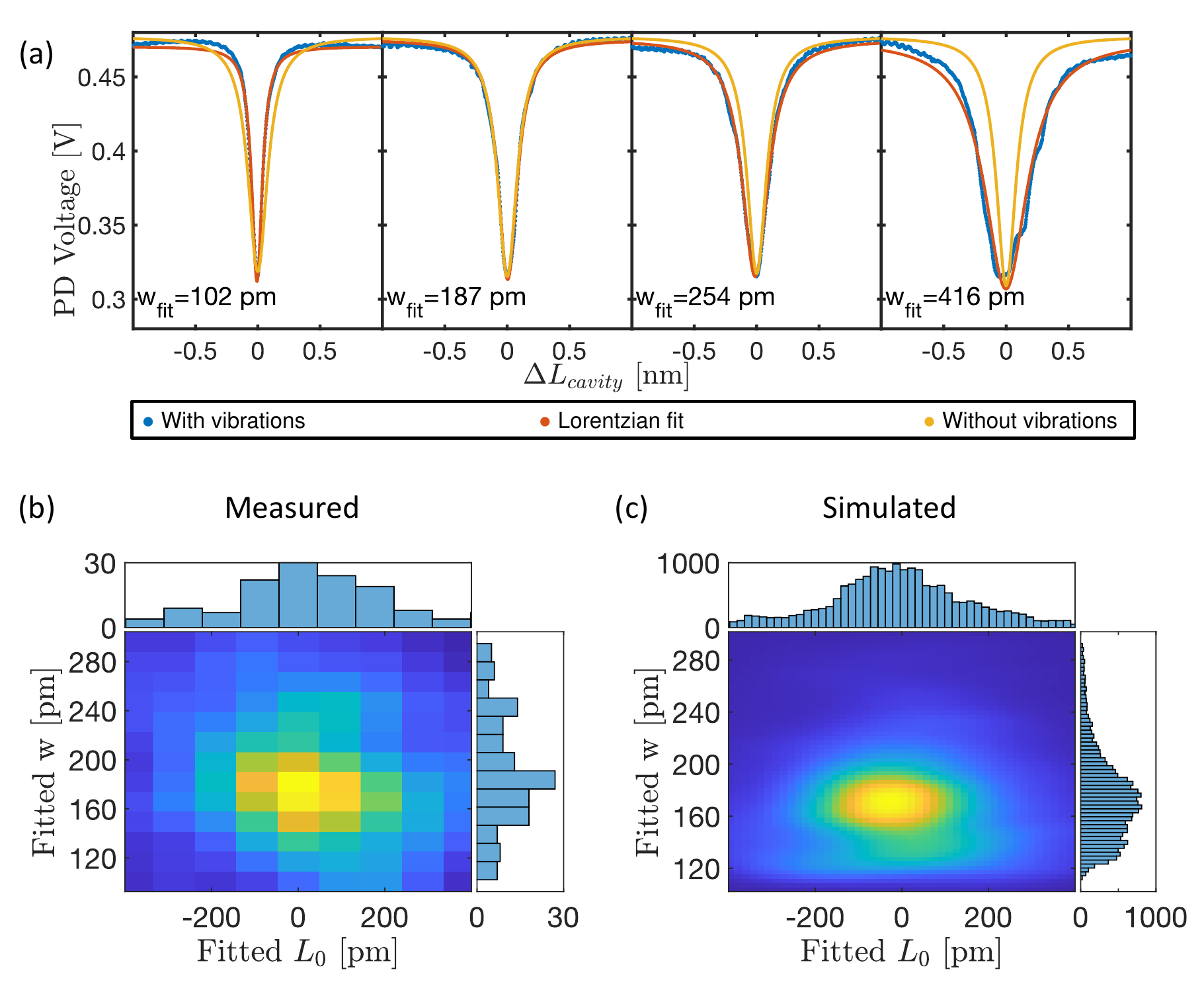}

\caption{\label{fig:scatter_histogram}(a) In blue four measurements of a cavity
resonance recorded at 4 K with cryocooler ON, with the same scanning
speed but at different times of the cryocooler cycle . In red the
corresponding Lorentzian fit, and in orange the real cavity resonance
measured at room temperature without vibrations. (b) Scatter plot
and relative histograms of the fitted widths and positions of the
cavity resonance, measured at 4 K. (c) Scatter plot and relative histograms
of the fitted widths and positions of the cavity resonance, result
of a computer simulation.}
\end{figure}

While this procedure takes into account the nonlinearity of the error
function, it still has some drawbacks: for the calibration we scan
the cavity length at a finite speed, and in presence of vibrations,
the shape of the reflection dip will change depending on the speed
at which the cavity is scanned, leading to shifts in the position
of the dip, broadening of the cavity reflection dip and asymmetries.
In Fig. \ref{fig:scatter_histogram}(a) we demonstrate this effect.
We have recorded four times a cavity dip, and we observe four very
different results, with extracted finesses varying from 1100 to 4600.
The real room-temperature finesse is measured to be F = 2500 at a
wavelength $\lambda$ = 935 nm. A simple way to deal with this is
to measure the width of the reflection dip at room temperature without
mechanical vibrations as a function of the wavelength of the laser,
and use this value to obtain the conversion factor of Eq. \ref{eq:Lorentzian_inverted},
as we expect the reflectivity of the mirrors to be constant from 300
K to 4 K.

As an additional proof and to have a better understanding of the phenomena,
we show here measurements of the shifts of the resonance widths and
dips ($w$, $L_{0}$) and compare it to simulations.
For the measurement we recorded 150 reflection dips at 4 K with the
mechanical noise generated by the cryocooler, by scanning repeatedly
the cavity length with the ring piezo. We fit each reflection dip
using Eq. \ref{eq:Lorentzian}, and we show the resulting scatter
plot in Fig. \ref{fig:scatter_histogram}(b). The data of the scatter
plot have been convoluted with a Gaussian mask function with witdth
of 10 pm, to smooth the data due to the low number of samples. The
mean value for the width is $w_{mean}=189$ pm, in very good agreement
with the real value of 185 pm. The mean value for $L_{0}$ is around
0 as we shifted the data manually, since we do not know the real position
of the resonance in absence of vibrations. For the simulation, we
used the power reflection coefficient of a Fabry-Perot cavity:

\begin{equation}
R(L)=\left|\frac{E_{reflected}}{E_{incident}}\right|^{2}=\left|\frac{r\left(\exp\left(i\frac{4\pi L}{\lambda}\right)-1\right)}{1-r^{2}\exp\left(i\frac{4\pi L}{\lambda}\right)}\right|^{2}\label{eq:reflection_dip}
\end{equation}

where $L=L_{piezo}+L_{noise}$. $L_{piezo}=v_{piezo}\times t$, and
$v_{piezo}$ is the scanning speed of the piezo. For $L_{noise}$
we used the measured displacements from Fig. \ref{fig:LowPower_vs_HighPower}
(a), recorded with a cavity locked on the side of the reflection dip,
where we compensated for the effect of the integrator: $L_{noise}=(1+I)\times L_{noise}^{lock}$.
We chose a mirror reflectivity R = $r^{2}$ = 0.988 which correspond
to a cavity with Finesse F $\sim2600$, and a resonance width of $w$
= 178 pm at the wavelength $\lambda$ = 935 nm. For all other free
parameters such as $v_{piezo}$ and the open loop gain of the integrator
$I$, we have used experimental values.

We have simulated 20000 reflection dips using Eq. \ref{eq:reflection_dip}
by randomly shifting in time the experimental $L_{noise}$, and we
fitted the curves with Eq. \ref{eq:Lorentzian} in order to extract
$w_{fit}$ and $L_{0}^{fit}$. The results are shown in Fig. \ref{fig:scatter_histogram}(c),
where the data have been smoothed by convolution with a Gaussian mask
function. The mean values are $w_{mean}=179$ pm and $L_{0}^{mean}=-1.7$
pm. Visually the distribution in Fig. \ref{fig:scatter_histogram}(c)
is clearly shifted towards negative $L_{0}$, we hypothesize that
this is because of the forces that keep together the two mirrors,
i.e. gravity and the three metal springs holding together the top
plate and the bottom part of the open cavity device. Since $L_{0}$
is the rest-length of the cavity, mechanical vibrations in the open-cavity
device will exert sinusoidal forces that act on the same direction
of gravity and the spring force for $L_{0}<0$, and in the opposite
direction for $L_{0}>0$, causing the asymmetry visible in the histogram.
The very good agreement between the measured data and the simulation,
as well as the comparison to the reflection widths measured in absence
of noise, shows us that indeed we can calibrate the photodiode signal
at 4 K with the value of width measured at room temperature for the
same wavelength.
\end{document}